\documentclass[fleqn,10pt]{wlscirep}
\usepackage[utf8]{inputenc}
\usepackage[T1]{fontenc}

\usepackage{multirow}

\title{Stimulation of soy seeds using environmentally friendly magnetic \& electric fields}

\author[1,*]{Agata Dziwulska-Hunek}
\author[2]{Agnieszka Niemczynowicz}
\author[3,4]{Rados\l{}aw A. Kycia}
\author[1]{Arkadiusz Matwijczuk}
\author[1]{Krzysztof Kornarzy\'{n}ski}
\author[5]{Joanna Stadnik}
\author[6]{Mariusz Szymanek}
\affil[1]{Department of Biophysics, University of Life Sciences in Lublin, Akademicka 13, 20-950; Lublin; Poland;}
\affil[2]{Department of Analysis and Differential Equations, University of Warmia and Mazury in Olsztyn, S\l{}oneczna 54, PL-10-710 Olsztyn, Poland;}
\affil[3]{Cracow University of Technology, Faculty of Computer Science and Telecommunications, Krakow, 31-155, Poland;}
\affil[4]{Department of Mathematics and Statistics, Masaryk Univeristy, Kotl\'{a}\v{r}sk\'{a} 267/2, 611 37 Brno, Czech Republic;}
\affil[5]{Department of Animal Material Technologies, University of Life Sciences in Lublin, Skromna 8, 20-704 Lublin; Poland;}
\affil[6]{Department of Agricultural, Forest and Transport Machinery, University of Life Sciences in Lublin, G\l{}eboka 28, 20-612 Lublin, Poland;}

\affil[*]{agata.dziwulska-hunek@up.lublin.pl}


\begin{abstract}
 The study analyzes the impact of constant and alternating magnetic fields and alternating electric fields on various growth parameters of soy plants: the germination energy and capacity, plants emergence and number, the Yield(II) of the fresh mass of seedlings, protein content, and photosynthetic parameters. Four cultivars were used: MAVKA, MERLIN, VIOLETTA, and ANUSZKA. Moreover, the advanced Machine Learning processing pipeline was proposed to distinguish the impact of physical factors on photosynthetic parameters. It is possible to distinguish exposition on different physical factors for the first three cultivars; therefore, it indicates that the EM factors have some observable effect on soy plants. Moreover, some influence of physical factors on growth parameters was observed. The use of ELM (Electromagnetic) fields had a positive impact on the germination rate in Merlin plants. The highest values were recorded for the constant magnetic field (CMF) - Merlin, and the lowest for the alternating electric field (AEF) - Violetta. An increase in terms of emergence and number of plants after seed stimulation was observed for the Mavka cultivar, except for the AEF treatment (number of plants after 30 days). In the case of the latter cultivar, a positive impact of ELM treatment was also observed in photosynthetic efficiency, electron transfer, and leaf greenness index (SPAD). In the case of the alternating magnetic field (AMF), increased leaf greenness index and fresh mass Yield(II) values were recorded in the Mavka cultivar, and a decrease thereof in Merlin plants. Moreover, CMF correlated positively with Yield(II), greenness index, and protein context in two of the studied cultivars - Mavka and Merlin. We used modern outlier detection and clustering methods for extracting classes that differentiate the various physical factors. The physical factors used to stimulate soy seeds are non-invasive and environmentally friendly as they introduce no changes to the natural agricultural ecosystem.
\end{abstract}
\begin{document}

\flushbottom
\maketitle

\section*{Introduction}

Soy is one of the most important arable crops cultivated worldwide \cite{1,2,3}. It is a leguminous plant whose symbiosis with Rhizobium bacteria enables it to bind nitrogen. This capacity makes for an inexpensive method of preserving soil fertility and improving plant yields \cite{4}. Soy seeds are a valuable source of protein ($30-40\%$) and fat ($19-20\%$), as well as minerals and vitamins, owing to which they play an essential part in human and animal nutrition \cite{5,6,7,8,9,10}. Soy is often used in food products and medicinal foodstuffs in the East \cite{5,11}. It is typically consumed in the form of sprouts, oil, paste \cite{12}, or soy meal \cite{13} soy milk \cite{14}.

Soy production is hindered by poor germination rates and low seed vitality. Therefore, it is necessary to employ methods facilitating better seed quality \cite{15}. In order to improve the germination rate, seeds were primed by Ancient Greeks by soaking them in milk or water as early as in 4th century BC \cite{16}. Literature provides reports on studies related to soaking seeds in hydrogen peroxide, potassium nitrate, or potassium chloride \cite{17}.
Moreover, stress caused by both abiotic (e.g., low or high temperature, water deficit or excess, salination, excessive sunlight, hail, etc.) and biotic factors (diseases, pests) has a considerable impact on plant growth, Yield(II), and seed storage \cite{18,19}.

Of the same, low temperatures are a major factor inhibiting the growth and development of plants, including soy. Soy germinates the most effectively in temperatures ranging from $10^oC$ to $30^oC$, with the rate increasing as the temperatures go up. Indeed, some authors claim that the optimum temperature is exactly $30^oC$ \cite{20,21,22,23,24,25}. Pre-soaking seeds further increases the germination rate in higher temperatures but decreases it in cooler conditions \cite{20}. In temperatures between $0^oC$ and $10^oC$, young soy plants tend to be damaged, and the stress has also been confirmed to impact photosynthetic processes negatively \cite{24}.

The seed moisture level is high during the harvest but decreases with long-term storage. This results in decreased seed quality, i.e., loss of vitality with the low vigor negatively influencing the germination rate, seedling growth, and plant regeneration. The conditions of correct cultivation depend on factors such as the vitality and age of seeds, conditions, and duration of storage, as well as the plant genotype. Moreover, reduced seed vitality hinders soy emergence, particularly in agriculture \cite{26}.

A key concern related to achieving higher efficiency in soy cultivation is the improvement of the germination rate, e.g., by employing one of the methods of processing seeds before sowing. The methods discussed in the following article include stimulation with alternating and constant magnetic fields and alternating electric fields \cite{27,28,29,30,31}. Due to their environmental friendliness, the techniques might potentially replace chemical treatment in the future. Other physical treatment methods have also been discussed in the literature, including laser light, gamma radiation, etc. \cite{18,32}. The pre-sowing treatment of seeds affects the physiological and biochemical properties of the material \cite{32}.  

Magnetic field stimulation was one of the broadly studied methods of seed processing analyzed in terms of impact on seed germination, plant growth, and biological parameters \cite{33, 34, 35, 36}. The technique remains the subject of considerable research scrutiny, mainly because of its non-invasiveness and environmentally friendly character, as well as its effectiveness in stimulating better seed germination and vigor \cite{37}. Research has also been conducted on the effects of an alternating magnetic field with the ICR cyclotron resonance frequency of Ca2+ ions on wheat seedlings, and the results evidenced a positive influence on the number of sprouts as well as the wet and dry weight of the seedlings \cite{38}. 

Stimulating seeds with an alternating electric field increased growth capacity in seven-year-old radish seeds by approx. $80\%$ \cite{39}, and the germination capacity of old tomato seeds (Halicz cultivar) by between $46\%$  and $80\%$ \cite{40} or even $100\%$ \cite{41}. 

From the technical perspective, chlorophyll fluorescence is measured using specialist fluorometers. The method can be used in studies on various aspects of photosynthesis in plant production, mainly to detect stress conditions but without identifying their exact nature \cite{42,43}. The method of determining leaf greenness using a portable chlorophyll meter (SPAD-502) can be used for fast and repeatable measurements in various plant species \cite{44,45}. 

However, global literature still lacks sufficient reports on the impact of soy seed stimulation using magnetic and electric fields on germination parameters, growth, development, Yield(II), or the efficiency of photosynthesis and the respective parameters of photosynthetic processes in leaves. Despite the ongoing scientific and technological advances, many secrets of the mechanisms related to plant growth are yet to be discovered. For this reason, we were driven to explore this research problem, which our group has been analyzing for over a dozen years. 

Moreover, the rising importance of advanced Machine Learning methods in science provides an excellent opportunity to use them in the research to discover non-trivial relationships in the data. The paper contains a well-thought clustering pipeline to check physical factors' influence on soy plants.

The paper is organized as follows: In the next section the methodology is outlined. Next, the results are presented. Finally, a discussion is given.

\section*{Results}

\subsection*{Parameters of germination and growth of soy plants}
\begin{figure}[ht]
\centering
\includegraphics[width=0.5\linewidth]{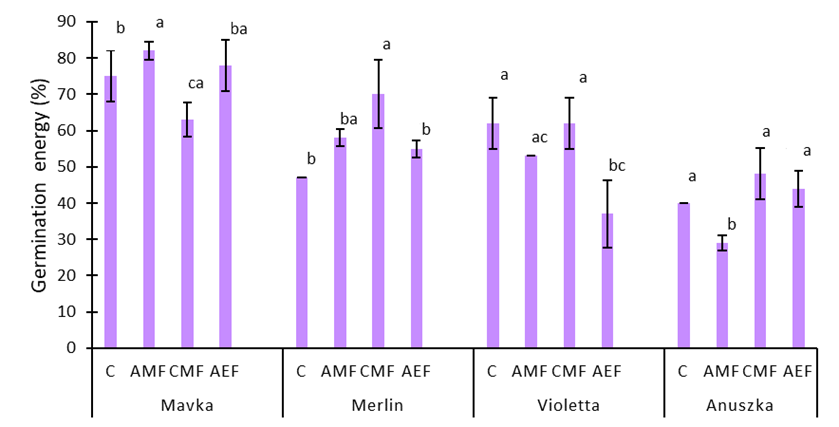}
\caption{Germination energy of soy seeds subjected to pre-sowing treatment using alternating electric field, constant and alternating magnetic field: C - control, AMF - alternating magnetic field, CMF - constant magnetic field, AEF - alternating electric field. Error bars - standard deviation. a-d - different letters indicate significant changes between the control and stimulated study groups, interaction between groups subjected to stimulation ($\alpha<0.05$). }
\label{fig:1}
\end{figure}

\begin{figure}[ht]
\centering
\includegraphics[width=0.5\linewidth]{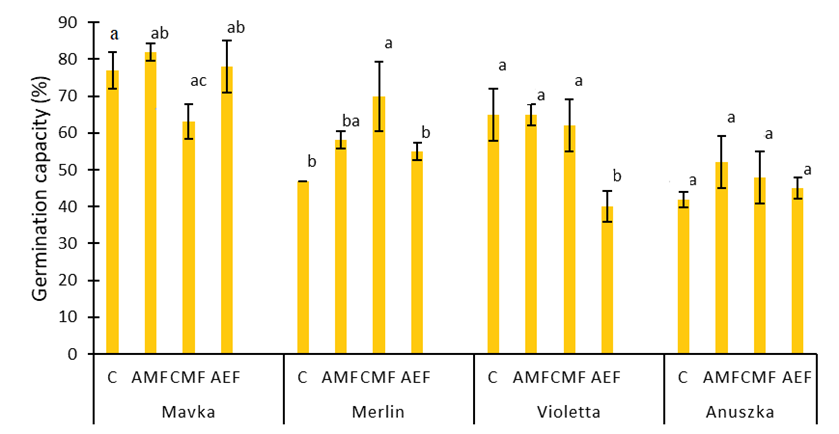}
\caption{Germination capacity of soy seeds subjected to pre-sowing treatment using alternating electric field, constant and alternating magnetic field: C - control, AMF - alternating magnetic field, CMF - constant magnetic field, AEF - alternating electric field. Error bars - standard deviation. a-d - different letters indicate significant changes between the control and stimulated study groups, interaction between groups subjected to stimulation ($\alpha<0.05$).}
\label{fig:2}
\end{figure}

The germination energy of soy seeds (Fig.\ref{fig:1}) was between $29\%$ (Anuszka AMF) and $82\%$ (Mavka AMF). In the range from $40\%$ (Violetta, AEF) to $82\%$ (Mavka, AMF), the germination capacity of the seeds changed as well (Fig. \ref{fig:2}). The significant, and simultaneously the highest, values of the germination parameters were observed for the Merlin and CMF cultivars ($23\%$), whereas the lowest extent of changes in Violetta and AEF plants ($25\%$), relative to the control (C).

\begin{figure}[ht]
\centering
\includegraphics[width=0.5\linewidth]{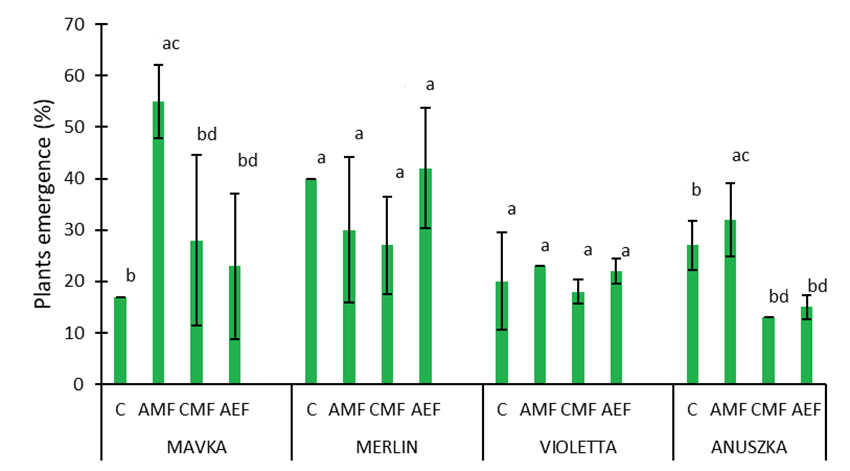}
\caption{Emergence of plants grown from soy seeds stimulated using alternating electric field, constant and alternating magnetic field: C - control, AMF - alternating magnetic field, CMF - constant magnetic field, AEF - alternating electric field. Error bars - standard deviation. a-d - different letters indicate significant changes between the control and stimulated study groups, interaction between groups subjected to stimulation  ($\alpha<0.05$).}
\label{fig:3}
\end{figure}

Relative to C, an increase in plant emergence was observed (Fig. \ref{fig:3}) in all the stimulated samples for the Mavka cultivar, with the highest increase of $38\%$ recorded for AMF. A significant $5\%$ increase was also observed in Anuszka plants, as well as a 105, but insignificant, decrease for Merlin plants. Under CMF stimulation, an insignificant $11\%$ increase was observed for the Mavka cultivar and a decrease by between $2$ and $14\%$ in the other three cultivars. The alternating magnetic field (AMF) triggered an increase in plant emergence, by between $2$ and $6\%$, in three of the cultivars and a decrease by $12\%$ in Anuszka plants. 

\begin{figure}[ht]
\centering
\includegraphics[width=0.5\linewidth]{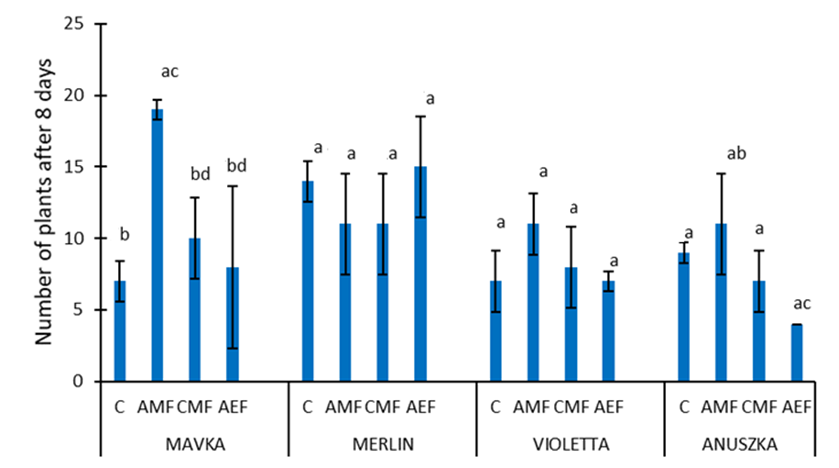}
\caption{Number of plants grown within 8 days from soy seeds stimulated using alternating electric field, constant and alternating magnetic field: C - control, AMF - alternating magnetic field, CMF - constant magnetic field, AEF - alternating electric field. Error bars - standard deviation. a-d - different letters indicate significant changes between the control and stimulated study groups, interaction between groups subjected to stimulation ($\alpha<0.05$).}
\label{fig:4}
\end{figure}

A nearly threefold increase in the number of plants was observed for the Mavka plants and AMF (Fig. \ref{fig:4}). A decrease by $56\%$ was recorded for Anuszka and AEF and by approx. $21\%$ for Merlin plants treated with AMF and CMF, otherwise no significant differences.  

\begin{figure}[ht]
\centering
\includegraphics[width=0.5\linewidth]{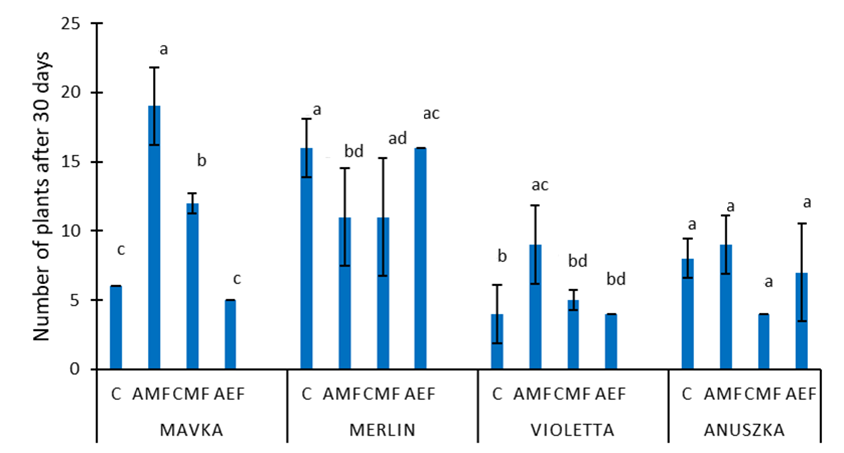}
\caption{Number of plants grown within 30 days from soy seeds stimulated using alternating electric field, constant and alternating magnetic field: C - control, AMF - alternating magnetic field, CMF - constant magnetic field, AEF - alternating electric field. Error bars - standard deviation. a-d - different letters indicate significant changes between the control and stimulated study groups, interaction between groups subjected to stimulation ($\alpha<0.05$).}
\label{fig:5}
\end{figure}

Similarly, after 30 days (Fig. \ref{fig:5}), the increase in the number of plants was the highest for Mavka (AMF). A twofold increase relative to the control was recorded for Mavka (CMF) and Violetta (AMF). A decrease by $31\%$ and $50\%$ was observed for Merlin (AMF and CMF) and Anuszka (CMF), respectively. 

The increase in the Yield(II) of fresh mass (Tab. \ref{tab:1}) was the highest (relative to C) for AMF, and the lowest for AEF in most of the studied cultivars. The highest recorded increase relative to C was $233\%$ (AMF, Violetta), $93\%$ (AEF, Violetta) and $71\%$ (CMF, Mavka). The lowest, $52\%$, was observed in Merlin plants (CMF).  

\begin{table}[]
\centering
\begin{tabular}{|l|llll|}
\hline
\multirow{3}{*}{\textbf{Variety}} & \multicolumn{4}{l|}{\begin{tabular}[c]{@{}l@{}}\textbf{Research}\\ \textbf{factors}\end{tabular}}                                                              \\ \cline{2-5} 
                         & \multicolumn{1}{l|}{\textbf{C}}                & \multicolumn{1}{l|}{\textbf{AMF}}               & \multicolumn{1}{l|}{\textbf{CMF}}               & \textbf{AEF}                \\ \cline{2-5} 
                         & \multicolumn{4}{l|}{\textbf{Yield(II) of fresh mass of seedlings after 30 days (g · pot$^-1$)}}                                                           \\ \hline
Mavka                    & \multicolumn{1}{l|}{$8.83 \pm 5.27$b} & \multicolumn{1}{l|}{$13.08 \pm 9.96$b} & \multicolumn{1}{l|}{$15.12\pm 6.75$ba} & $7.11\pm 4.88$bc   \\ \hline
Merlin                   & \multicolumn{1}{l|}{$19.44\pm 4.17$a} & \multicolumn{1}{l|}{$13.29\pm 3.41$ab} & \multicolumn{1}{l|}{$9.30\pm 8.22$b}   & $12.69 \pm 4.41$ab \\ \hline
Violetta                 & \multicolumn{1}{l|}{$2.30\pm 2.29$a}  & \multicolumn{1}{l|}{$5.36 \pm 4.45$a}  & \multicolumn{1}{l|}{$2.73\pm 0.94$a}   & $4.43\pm 2.12$a    \\ \hline
Anuszka                  & \multicolumn{1}{l|}{$5.00\pm 2.78$a}  & \multicolumn{1}{l|}{$6.29 \pm 4.38$a}  & \multicolumn{1}{l|}{$4.50 \pm 2.37$a}  & $3.21\pm 2.96$a    \\ \hline
\end{tabular}
\caption{C - control, AMF - alternating magnetic field, CMF - constant magnetic field, AEF - alternating electric field.
$\pm$ standard deviation , mark a-d in rows mean statistical differences between control and research objects, interaction between groups subjected to simulation ($\alpha<0.05$). }
\label{tab:1}
\end{table}

\subsection*{Relative correlations between plant germination and growth parameters}
\begin{figure}[ht]
\centering
\includegraphics[width=0.5\linewidth]{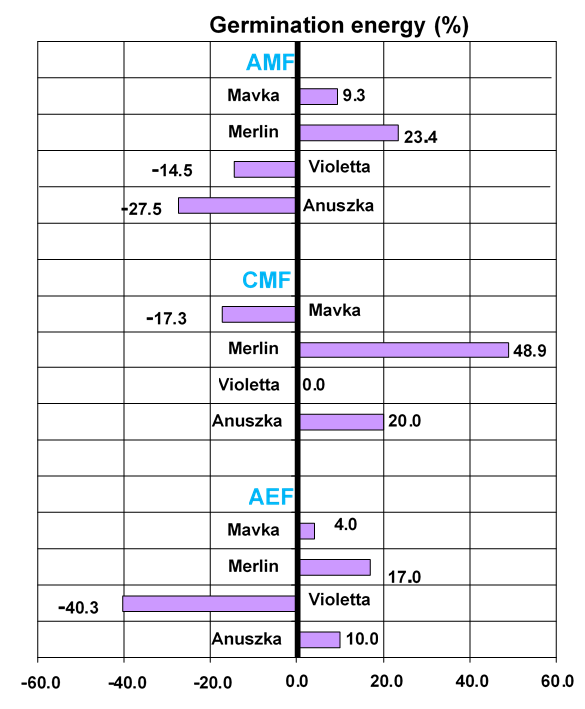}
\caption{Graphic representation of the relative germination energy values C - control, AMF - alternating magnetic field, CMF - constant magnetic field, AEF - alternating electric field.}
\label{fig:6}
\end{figure}

After AMF stimulation, a positive impact on the germination energy (Fig. \ref{fig:6}) was observed for Merlin ($23,4\%$) and Mavka ($9.3\%$) plants, and a negative impact for Violetta ($14.5\%$) and Anuszka ($27.5\%$) cultivars. After CMF stimulation, a positive impact on the germination energy was recorded for Merlin ($48.9\%$) and Anuszka ($20\%$), and a negative impact for Mavka ($17.3\%$) plants.

\begin{figure}[ht]
\centering
\includegraphics[width=0.5\linewidth]{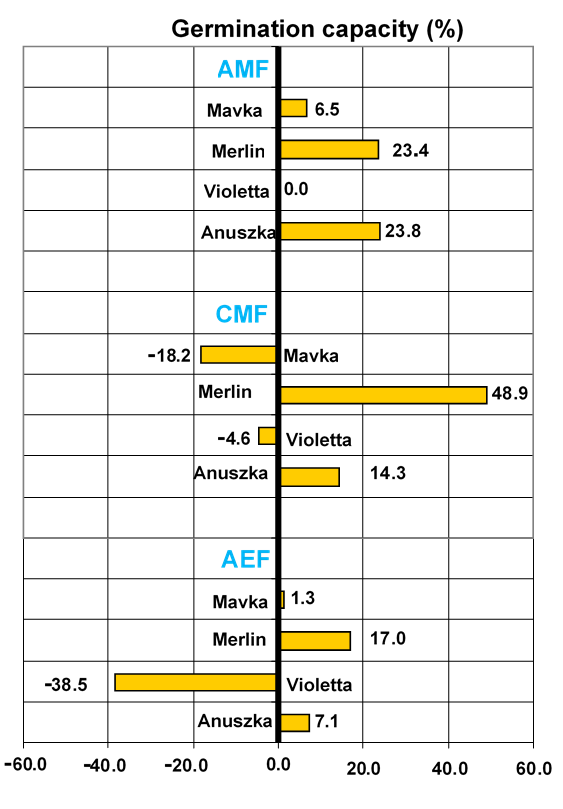}
\caption{Graphic representation of the relative germination capacity values C - control, AMF - alternating magnetic field, CMF - constant magnetic field, AEF - alternating electric field.}
\label{fig:7}
\end{figure}

Meanwhile, the AEF field had a positive impact on the germination energy and capacity of Merlin ($17\%$), Anuszka ($10\%$), and Mavka ($4\%$) plants, whereas a negative impact on Violetta plants ($40.3\%$).
The positive impact of AMF on the germination capacity (Fig. \ref{fig:7}) was observed for the  Mavka ($6.5\%$), Merlin ($23.4\%$), and Anuszka ($23.8\%$) cultivars, but no effects were recorded for Violetta plants. In turn, CMF positively influenced Merlin ($48.9\%$) and Anuszka ($14.3\%$) but had a negative impact on Mavka ($18.2\%$) and Violetta ($4.6\%$) plants. In the case of AEF, positive effects were observed for the Mavka ($1.3\%$), Anuszka ($7.1\%$), and Merlin ($17.0\%$) cultivars, whereas negative only for the Violetta ($38.5\%$) cultivar.

\begin{figure}[ht]
\centering
\includegraphics[width=0.5\linewidth]{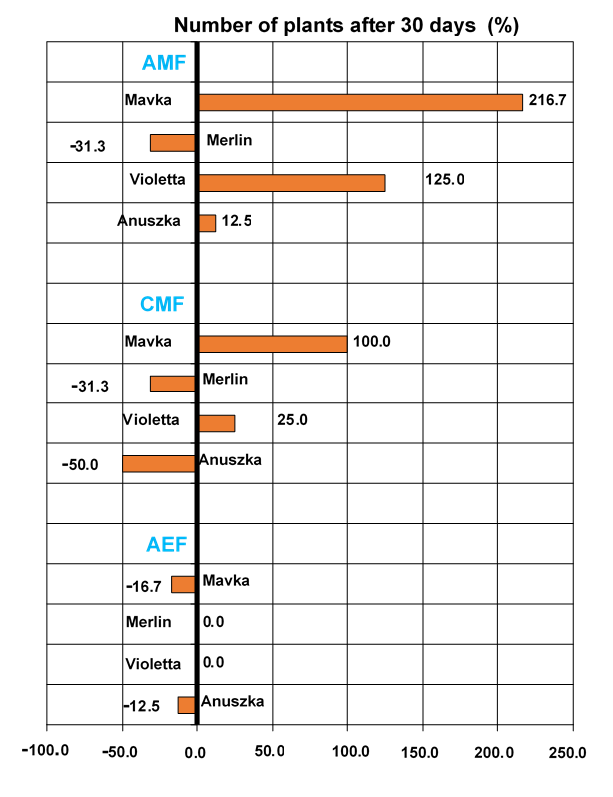}
\caption{Graphic representation of the relative plant numbers C - control, AMF - alternating magnetic field, CMF - constant magnetic field, AEF - alternating electric field.}
\label{fig:8}
\end{figure}

In terms of the number of plants after 30 days (Fig. \ref{fig:8}), a very high increase ($216.7\%$) was recorded for AMF in the case of Mavka plants, as well as a significant ($125\%$) increase for the Violetta cultivar. Anuszka plants also increased, albeit considerably lower ($12.5\%$). The negative impact of AMF on the number of plants was recorded only for Merlin plants ($31.3\%$). As evidenced by the above, in certain situations the stimulating factors can also have a negative impact. For CMF, positive results were observed for Mavka ($100\%$) and Violetta ($25\%$) plants, but at the same time negative results were returned for Merlin ($31.3\%$) and Anuszka ($50\%$) cultivars.

\subsection*{Leaf protein content }
\begin{figure}[ht]
\centering
\includegraphics[width=0.5\linewidth]{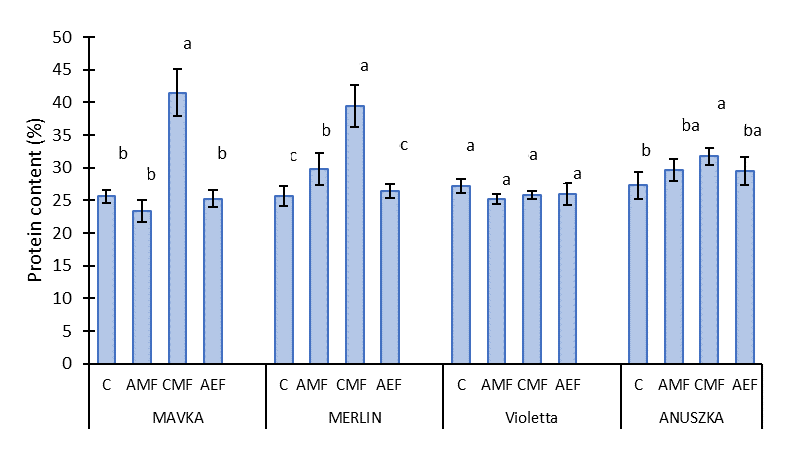}
\caption{Representation of the relative plant numbers C - control, AMF - alternating magnetic field, CMF - constant magnetic field, AEF - alternating electric field.Protein content in soybean leaves C - control, AMF - alternating magnetic field, CMF - constant magnetic field, AEF - alternating electric field. Error bars - standard deviation. a-d - different letters indicate significant changes between the control and stimulated study groups, interaction between groups subjected to stimulation ($\alpha<0.05$).}
\label{fig:9}
\end{figure}

After the field stimulation, the protein content (Fig. \ref{fig:9}) in half of the analyzed soy cultivars was increased compared to the control. The highest values were observed in two cultivars after AMF stimulation, with a significant increase by $13\%$ (Merlin) and $16\%$ (Mavka) relative to C. It is also noteworthy that the stimulation increased the protein content in Anuszka plants by $3-5\%$. The least amounts of protein were observed in the AMF sample ($2-3\%$) but the differences were statistically insignificant.

\subsection*{Relative correlations between photosynthetic parameters in soy leaves}

\begin{figure}[ht]
\centering
\includegraphics[width=0.5\linewidth]{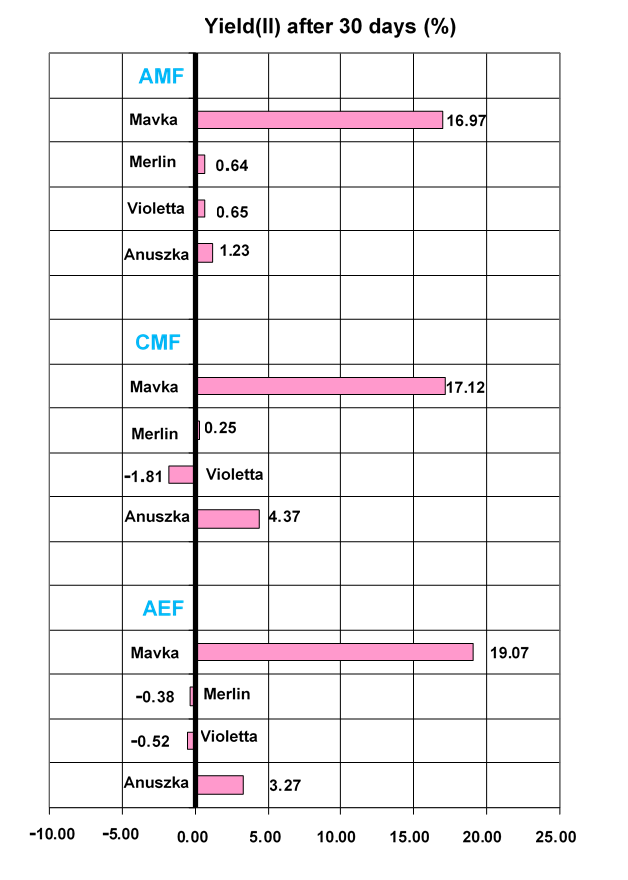}
\caption{Representation of relative Yield(II) values, C - control, AMF - alternating magnetic field, CMF - constant magnetic field, AEF - alternating electric field. }
\label{fig:9_10}
\end{figure}

\begin{figure}[ht]
\centering
\includegraphics[width=0.5\linewidth]{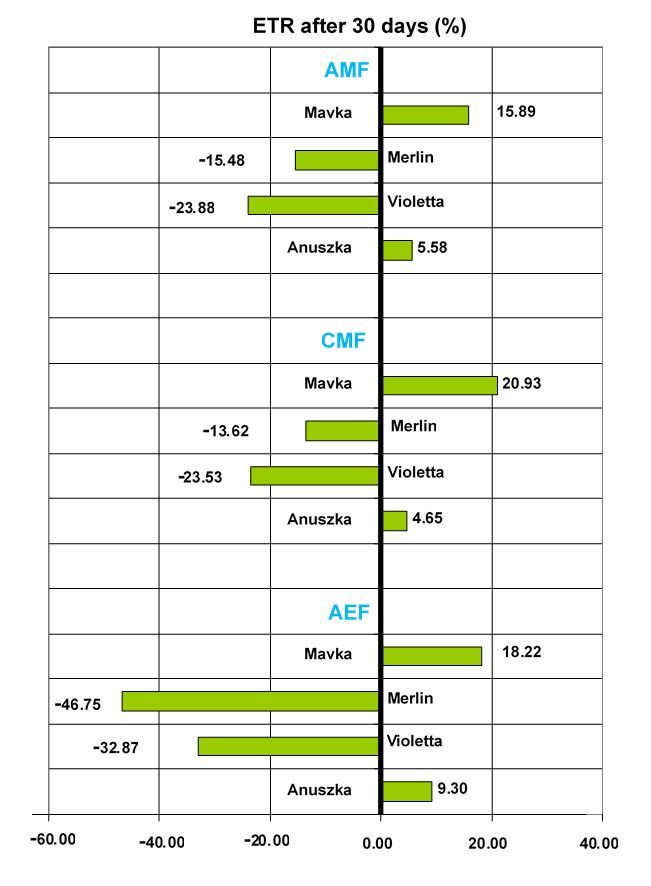}
\caption{Representation of relative ETR values, C - control, AMF - alternating magnetic field, CMF - constant magnetic field, AEF - alternating electric field.}
\label{fig:10_11}
\end{figure}

\begin{figure}[ht]
\centering
\includegraphics[width=0.5\linewidth]{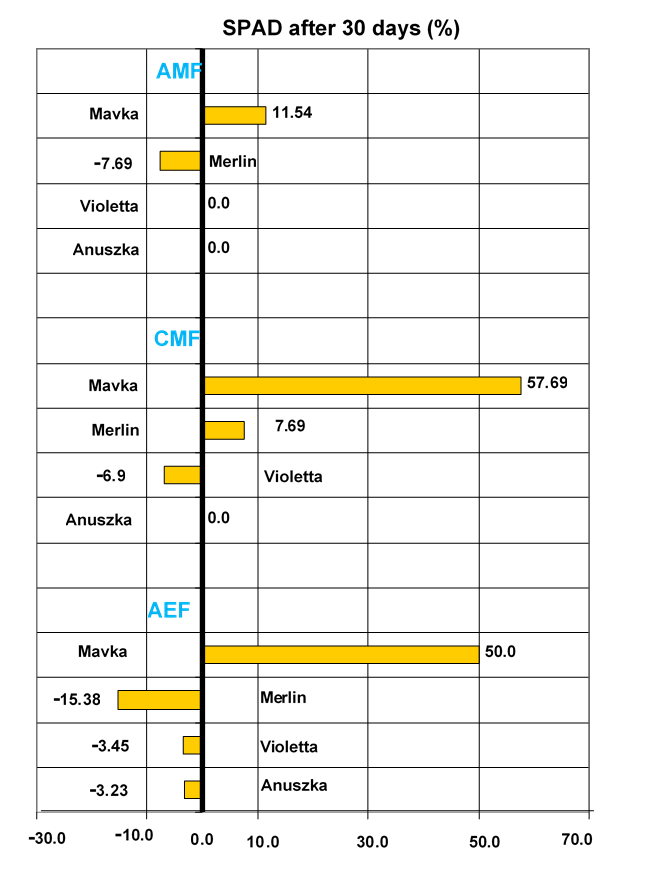}
\caption{Representation of relative SPAD greenness index values, C - control, AMF - alternating magnetic field, CMF - constant magnetic field, AEF - alternating electric field.}
\label{fig:11_12}
\end{figure}

In the case of photosynthetic parameters (Figs. \ref{fig:9_10}, \ref{fig:10_11}, \ref{fig:11_12} ), in order to determine the impact of the stimulation, we sought correlations between the respective cultivars and the physical stimuli used, to identify similarities and repeatability. 

The observed influence of ELM stimulation on the photosynthetic index (Fig. \ref{fig:9_10}) was both positive and negative compared to the control sample. In this respect, a clearly positive impact of seed treatment was recorded for two of the cultivars (Mavka and Anuszka). The highest value of Yield(II) was obtained for AEF stimulation (Mavka, (increase by $19\%$), and the lowest for CMF (Violetta, decrease by approx. $2\%$). 

Electron transfer (Fig. \ref{fig:10_11}) was positively influenced by all the fields in Mavka and Anuszka plants, and negatively in the other two cultivars (Merlin and Violetta).   The highest value relative to the control was observed for CMF ($21\%$ increase, Mavka) and the lowest for AEF ($47\%$ decrease, Merlin). 

The impact of field treatment on the SPAD greenness index (Fig. \ref{fig:11_12}) was positive in one cultivar (Mavka) and negative in the other three. The highest increase was observed for CMF ($58\%$, Mavka) and the highest decrease for AEF (over $15\%$, Merlin).
A positive impact of alternating magnetic field treatment was identified for the Yield(II) and ETR correlation in the Mavka and Anuszka cultivars; the other two cultivars showed negative effects in this respect. Positive effects of ELM fields in the correlation of three parameters: Yield(II), ETR, and SPAD could be observed only in one cultivar (Mavka). 

A negative impact of alternating electric fields was observed in two of the studied cultivars (Merlin and Violetta) and of constant electric fields in one (Violetta), in a correlation between Yield(II), ETR, and SPAD. 

To recapitulate, in the study on using ELM fields to treat soy seeds, the results related to Yield(II), ETR, and SPAD were inconsistent and evidenced both positive and negative effects of the treatment. One should note, in particular, the object where the use of CMF in combination with the Mavka cultivar produced a clearly positive result in terms of protein content and photosynthetic parameters (photosynthesis, electron transfer, greenness index).

\subsection*{Clustering analysis}
The previous sections indicate a possible impact on averaged values of soy parameters. Therefore, it is essential to cross-check these relations for individual samples.

Our first attempt to distinguish the impact of physical factors on soy plants was to use supervised learning for the whole range of data (not averaged). In order to maximize the number of records, only four features were selected: Leaf Greenning index after 15 and 30 days and Yield(II) after 15 and 30 days. The data labels were selected as either varieties, electromagnetic factors, or both. We used multiple classes or a one-class-against-others approach with proper weighting. The standard pipeline was applied that consists of the following steps: 
\begin{enumerate}
    \item Standard scaling of data.
    \item Dimensional reduction of data in terms of Principal Component Analysis.
    \item Cross validate models.
\end{enumerate}
with models including Naive Bayesian, Decision Tree, Support Vector Machine, and Neural Network. Unfortunately, averaged weighted accuracy and other metrics suggest that the effectiveness of prediction is at the level of about $50\%$.

Due to the ineffectiveness of this approach, we proposed another method that can distinguish various classes based on unsupervised learning (clustering) and nonlinear transformation connected with dimensionality reduction. The redesigned analysis pipeline consists of several steps: 
\begin{enumerate}
    \item Remove outliers. 
    \item Normalize data. 
    \item Reduce data dimensionality and nonlinear transform/projection to the three-dimensional space.
    \item Compute the optimal number of clusters using the averaged silhouette score.
    \item Clustering.
\end{enumerate}

\subsubsection*{Designing of clustering algorithm}
Since the classical machine learning approach was not conclusive, the new, more robust approach was designed. For the analysis, the algorithms from Scikit Learn 0.22.2.post1 version of Python library \cite{46} was used. 

It was checked that the following procedure provides the best uniform clusters for the following parameters: 
\begin{itemize}
    \item {SPAD greenness index after 15 and 30 days}
    \item {Yield(II) after 15 and 30 days}
    \item {ETR after 15 and 30 days}
\end{itemize}

The first step is to clean data from the outliers. This step is necessary since, by visual inspection, it was spotted that some of the values look unreliable. The following reasoning explained this: some complicated measurements such as ETR or Yield(II) heavily rely on multiple factors such as sample preparation and proper measurement procedure that is error susceptible. Therefore errors should occur when a high number of measurements is produced. The two standard techniques for outlier/anomaly detection were checked:
\begin{itemize}
    \item {Isolation Forest - sensitive to global outliers}
    \item {Local Outliers Factory - isolates local outliers}
\end{itemize}
Both algorithms have their advantages and drawbacks \cite{IsolationForestVSLocalOutlierFactory}. In order to restrict the globality of the Isolation Forest, the run was performed on a specific cultivar, however, it was checked that although Isolation Forest removes more possible outliers, it produces at further steps more uniform clusters. In most cases the outliers detected by the Local Outliers factory are included in the outliers found by the Isolation Forest algorithm.

In the next step the Hopkins statistics was used to check clustering tendency in the data. The value is $0.8784$, which indicates a high tendency to cluster. 

Then for each object, standardization to the mean $0$ and STD $1$ was made. Since it was checked that the clusters do not separate well upon linear transformation through PCA, it was decided to perform nonlinear transformation with a dimensional reduction to $3$ dimensions using UMAP (Uniform Manifold Approximation and Projection for Dimension Reduction) \cite{UMAP_paper, UMAP_Webpage}. This algorithm has similar or better performance for many cases than the other state-of-the-art nonlinear transformation and dimensionality reduction algorithm t-SNE used commonly for clustering \cite{UMAP_paper}.

Finally, the optimal number of clusters was extracted using two independent techniques: the first one is the maximization of averaged silhouette score that also provides the information on the balance of clusters; the second one is the elbow rule on the SSE vs. the number of clusters plot. After this check, the standard k-Means clustering algorithm was used to identify clusters and check their content.

The result of the procedure is presented in the following subsection.

\subsubsection*{Clustering of cultivars}
\begin{figure}[htb]
    \centering
    \includegraphics[width=0.9\textwidth]{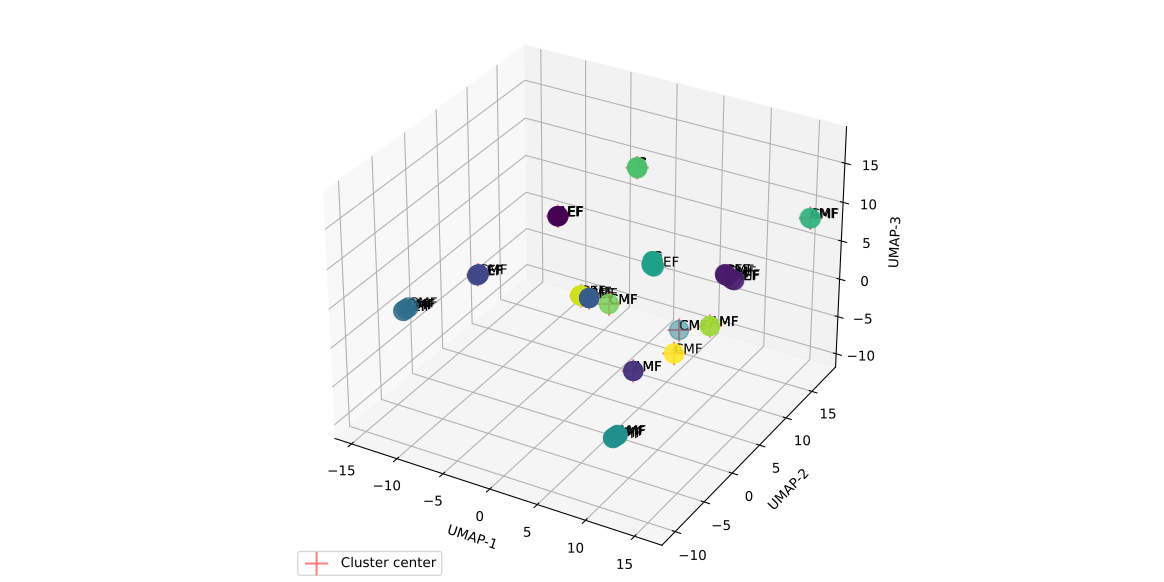}
    \caption{15 clusters for MAVKA}
    \label{fig:MAVKA15clusters}
\end{figure}

\textbf{MAVKA:} This class contains $51$ records. After scaling and UMAP transformation and projection the averaged silhouette score was optimized for 15 clusters. The k-Means clustering is presented in Fig. \ref{fig:MAVKA15clusters}. 
and the clusters are presented in Tab. \ref{tab:MAVKAclusters}. It is visible that most of the clusters are clean and therefore our approach can distinguish seeds exposed by different physical factors.
\begin{table}[]
\centering
\begin{tabular}{|l|l|}
\hline
\textbf{Cluster number} & \textbf{Physical factor}                     \\ \hline
0              & AEF, AEF, AEF, AEF                 \\ \hline
1              & AEF, AEF, AEF, C, C, CMF, CMF, CMF \\ \hline
2              & AMF, AMF                           \\ \hline
3              & AEF, AEF, CMF                      \\ \hline
4              & AEF, CMF                           \\ \hline
5              & AEF, CMF, CMF, CMF, CMF            \\ \hline
6              & CMF, CMF                           \\ \hline
7              & AMF, AMF, AMF, AMF                 \\ \hline
8              & AEF, C, C, C, C, C                 \\ \hline
9              & AMF, AMF, C                        \\ \hline
10             & C, C, C                            \\ \hline
11             & CMF, CMF                           \\ \hline
12             & AMF, AMF                           \\ \hline
13             & AEF, AMF, C                        \\ \hline
14             & AMF, C             \\ \hline
\end{tabular}
\caption{Content of the clusters for MAVKA cultivar.}
\label{tab:MAVKAclusters}
\end{table}
\\
\textbf{MERLIN:} This class contains $47$ samples. The similar procedure as above provided $13$ clusters with the content in Tab. \ref{tab:MERLINclusters}. The obtained clusters are also distinguished.
\begin{table}[]
\centering
\begin{tabular}{|l|l|}
\hline
\textbf{Cluster number} & \textbf{Physical factor}                            \\ \hline
0              & AEF, AEF, AEF, AMF                        \\ \hline
1              & AEF, AEF, AEF, AEF                        \\ \hline
2              & C, C, C, C, C, C                          \\ \hline
3              & AEF, C, CMF, CMF, CMF, CMF, CMF, CMF, CMF \\ \hline
4              & AMF, AMF, AMF                             \\ \hline
5              & AEF, AEF, AEF                             \\ \hline
6              & C, C                                      \\ \hline
7              & AEF, AMF, AMF, AMF                        \\ \hline
8              & C, C                                      \\ \hline
9              & AMF, AMF, AMF                             \\ \hline
10             & AMF, C, C                                 \\ \hline
11             & CMF, CMF                                  \\ \hline
12             & AMF, AMF                                  \\ \hline
\end{tabular}
\caption{Content of the clusters for MERLIN cultivar.}
\label{tab:MERLINclusters}
\end{table}
\\
\textbf{VIOLETTA:} The sample contains $48$ records. After processing the cluster content is presented in Tab. \ref{tab:VIOLETTAclusters}.
\begin{table}[]
\centering
\begin{tabular}{|l|l|}
\hline
\textbf{Cluster number} & \textbf{Physical factor}                                                                             \\ \hline
0              & \begin{tabular}[c]{@{}l@{}}AEF, AEF, AEF, AEF, AEF, AEF, AEF, CMF,\\ CMF, CMF\end{tabular} \\ \hline
1              & AMF, AMF, AMF, AMF, C, C, CMF, CMF                                                         \\ \hline
2              & C, C, C, C, C, CMF                                                                         \\ \hline
3              & AMF, AMF, AMF, AMF, CMF                                                                    \\ \hline
4              & AEF, AEF, AEF, C, CMF                                                                      \\ \hline
5              & AEF, C, C, CMF                                                                             \\ \hline
6              & CMF, CMF, CMF                                                                              \\ \hline
7              & AMF, C, CMF                                                                                \\ \hline
8              & AMF, AMF                                                                                   \\ \hline
9              & AEF, CMF                                                                                   \\ \hline
\end{tabular}
\caption{Content of the clusters for VIOLETTA cultivar.}
\label{tab:VIOLETTAclusters}
\end{table}
In this case some clusters are mixed with no distinguished leading factor, e.g., clusters no. 7 and 9.
\\
\textbf{ANUSZKA:} For this cultivar there are $46$ records, and the optimal number of clusters is determined to be $10$. The content of clusters is presented in Tab. \ref{tab:ANUSZKAclusters}. In this case the clusters are the most nonuniform. The proposed method can distinguish some clear clusters, but most are nonuniform. 

\begin{table}[]
\centering
\begin{tabular}{|l|l|}
\hline
\textbf{Cluster number} & \textbf{Physical factor}                           \\ \hline
0              & AEF, AEF, AMF                            \\ \hline
1              & AEF, CMF                                 \\ \hline
2              & AMF, AMF, AMF                            \\ \hline
3              & AEF, AEF, AEF, AMF, C, C, CMF, CMF       \\ \hline
4              & AMF, AMF, C, CMF                         \\ \hline
5              & AMF, AMF, C, C, C, C, CMF, CMF, CMF, CMF \\ \hline
6              & AMF, AMF, AMF                            \\ \hline
7              & AEF, AEF, AEF, AEF, AMF, C, C, CMF       \\ \hline
8              & AMF, C, CMF                              \\ \hline
9              & C, C                                     \\ \hline
\end{tabular}
\caption{Content of the clusters for ANUSZKA cultivar.}
\label{tab:ANUSZKAclusters}
\end{table}

\subsubsection*{Discussion}

The method to distinguish plants under various physical factors was proposed and checked on specific examples. The first two cultivars (MAVKA, MERLIN) provide relatively high predictive power, indicating that the physical factors alter Yield(II), ETR, and leaf greenness parameters. This indicates a potential field to explore in more detailed experiments with higher statistics. VIOLETTA shows a smaller distinction under EM factors. The distinction is smaller for ANUSZKA cultivar. This can suggest that ANUSZKA is less susceptible to the used factors.

The proposed processing/clustering pipeline can be used in more advanced studies of the influence of external factors on plants growth. Moreover, the failure of standard supervised ML methods indicates that the influence of EM factors can be masked by some nonlinear relations that more advanced transformations like UMAP can unmask.

\section*{Conclusions}
A positive impact of ELM fields on germination energy and capacity was observed for the Merlin cultivar, for which the significant (and highest) increase reached $23\%$ (CMF). At the same time, in the case of Violetta plants, a significant, approx. $40\%$ decrease was observed in the AMF sample. Emergence and plant numbers increased in all fields sown with Mavka seeds except the electric sample (number of plants after 30 days). The value of the SPAD greenness index in the correlation of 30-day plants to 15-day plants increased relative to the control for most of the analyzed cultivars and respective ELM fields. Howerer, the SPAD value as such for 15- and 30-day plants decreased (or remained unchanged) for most cultivars and stimulation treatments. This may evidence a negative effect of ELM fields on the chlorophyll content, making them a stressogenic factor in this respect. 

In the case of protein content in soy plants, in most cultivar and stimulant combinations was observed to increase (a positive effect relative to the control). The most significant results were noted under CMF stimulation in Mavka and Merlin plants. 

Using ELM stimulation on soy seeds positively influenced the photosynthetic efficiency, electron transport rate, and SPAD chlorophyll content in Mavka plants. The results were less decisive in the remaining three cultivars, showing both positive and negative impacts. In Merlin plants, a decrease in the greenness index and plant number after 30 days was recorded for the AMF treatment. 

To recapitulate, it can be concluded that changes in the parameters of germination and plant growth, as well as the chlorophyll and protein content, depended both on the particular ELM treatment employed and the soy cultivar in question. It is noteworthy that physical factors may indeed be used to stimulate soy seeds in a non-invasive and environmentally friendly manner as they introduce no changes to the natural agricultural ecosystem. Stimulation can improve germination, Yield(II), and photosynthetic efficiency, but further research in this context is definitely needed. 

Moreover, we proposed a clustering algorithm that can be used to examine the nonlinear relation between external stimuli on plants. It is based on the UMAP method. 

\section*{Materials and methods}
\subsection*{Plants}
The research material comprised seeds of four soy cultivars: Mavka, Merlin, Violetta, and Anuszka, harvested in 2017 and stored under room conditions. In 2018, an experiment was conducted at the Chair of Biophysics of the University of Life Sciences in Lublin. It entailed two stages: determination of the germination energy and capacity of seeds placed on Petri dishes, and determination of the rate of emergence, number of plants, fresh mass Yield(II), photosynthetic parameters, and leaf protein content in pot-grown plants.  

We confirm that experimental  studies on used plants cultivated in the study, are comply with relevant institutional and national guidelines and legislation effectual at Research Centre for Cultivar Testing \cite{Zgoda}.

\subsection*{Pre-sowing stimulation of soy seeds}
Before sowing, soy seeds were subjected to stimulation using an alternating magnetic field with the induction of $B = 30 mT$ and exposure time of $t = 60 s$, a constant magnetic field with the induction of $B = 130 mT$ and exposure time $t = 17 h$, or an alternating electric field with the intensity of $E = 5 kV cm^{-1}$ for $t = 60 s$, respectively designated as: AMF, CMF, and AEF as well as C - control sample (non-stimulated seeds).

\subsection*{Laboratory experiment}
The germination energy (GE) and capacity (GC) were determined in accordance with the applicable ISTA procedure \cite{80} after $5$ and $8$ days of the experiment. 
\begin{equation}
    GE(\%)=\frac{N_{5}}{N}100\%,
\end{equation}
\begin{equation}
    GC(\%)=\frac{N_{8}}{N}100\%.
\end{equation}
where GE - germination energy, GC - germination capacity, N - number of seeds sown on the dish (units), $N_5$ - number of seeds germinating after 5 days (units), $N_8$ - number of seeds germinating after 8 days (units). 

After stimulation, the soy seeds were sown in 30 parts in each pot ($1 dm^3$) filled with universal soil, in $4$ replicates each treatment. Seed sowing depth was $2 cm$ and covered with universal soil for $1 cm$. Soil humidity was maintained by ensuring constant availability of water by placing the pots in horticultural trays filled with water of constant volume throughout the experiment duration. Samples were placed inside a climate chamber, under day/night 16/8 h illumination, day/night temperatures of $23^{o}C / 12^{o}C(\pm 2^{o}C)$, and $E = 600 lx$ light intensity. 

The experimental measurements of each treatment are presented as mean plant emergence (after 5 days), the number of plants (after 8 days and 30 days), the Yield(II) of the fresh mass of seedlings (after 30 days), SPAD greenness index, quantum photochemical Yield(II), ETR - electron transport rate, and protein content in leaves. The PE (plant emergency) is defined as
\begin{equation}
    PE(\%)=\frac{N_{5}}{N}100\%,
\end{equation}
where N - number of seeds sown in pots (units), N5 - number of seeds germinating after 5 days (units).

\subsection*{Photosynthetic parameters}
The greenness index, photosynthetic efficiency (Yield(II), and electron transport rate (ETR) were determined after 15 and 30 days for 16 randomly selected leaves, randomly selected from the 4 replications in each treatment. The greenness index was measured with the use of a SPAD-502 Chlorophyll Meter \cite{45, 46}. Yield(II) and electron transport rate were measured using a MINI-PAM 2000 WALTZ Photosynthesis Yield Analyzer (Germany). 

The following formulas were used: 
\begin{equation}
    YIELD(II)=\frac{Y}{1000}=\frac{M-F}{M},
\end{equation}
where Yield(II) - current quantum efficiency PSII, M - Maximal fluorescence Yield(II) (M = Fm or Fm') measured during the last saturating light pulse triggered by START, F - Momentary fluorescence Yield(II) displaying small fluctuations.

Display of current factor applied for calculation of relative electron transport rate (ETR), which for a standard leaf is defined as follows:
\begin{equation}
    ETR = YIELD(II) \times PAR \times 0.5 \times 0.84,
\end{equation}
where ETR - relative electron transport rate, PAR - Photosynthetically active radiation. The standard factor 0.84 corresponds to the fraction of incident light absorbed by a leaf. The preset value, which corresponds to an average observed with various leaf species, can be modified via SET and the arrow-keys.

\subsection*{Kjeldahl's method}
The protein content in leaves was determined with Kjeldahl's method, following PN-EN\_ISO 5983-1:2006/AC:2009 \cite{81}, in triplicate. Soy leaves were collected after 30 days. Each sample was weighed to obtain $1 g$ of fresh mass. The method entails mineralization of the sample, the distillation of ammonia, and titration of the ammonia released. This allows the nitrogen content to be calculated under the protein formula. Each study sample was repeated in triplicate. 

\subsection*{Statistical analysis}
The obtained results were processed by way of ANOVA variance analysis, LSD test at the significance level of $\alpha < 0.05$, using STATISTICA 13.1 software. Significance was determined between the research factors for the respective cultivars and the control (C).
Moreover, the research results were processed and presented using an innovative, proprietary method for determining the percentage of relative relation compared to the control. 

The percentage of relative relation was calculated using:
\begin{equation}
    X(\%)=\left(\frac{A}{C}-1\right)100\%,
\end{equation}
where A - value obtained, C - control value. Positive results indicate  the percentage of relative increase, negative values - relative decrease, compared to the control. 

\subsection*{Machine Learning - clustering}
We used Python 3.8.5 interpreter and Machine Learning library: Scikit-Learn 0.22.2 \cite{46}. For UMAP analysis we used reference implementation \cite{UMAP_Webpage}.



\section*{Acknowledgements}
This research was funded by The University of Life Sciences in Lublin. R.K. part of the research was supported by the GACR grant GA22-00091S, the grant 8J20DE004 of the Ministry of Education, Youth and Sports of the CR, and Masaryk University grant MUNI/A/1092/2021. AN and RK also thank the SyMat COST Action (CA18223) for partial support.

\section*{Author contributions statement}
Conceptualization, A.D.-H., K.K. A.N. and R.A.K.; Data curation, A.D.- H., K.K., J.S., A.N, R.A. K. and M. S.; Formal analysis, A.D.-H., K.K., A.M., A.N., R.A.K. and M.S.; Funding acquisition, A.M.; Investigation, A.D.-H.; Methodology, A.D.-H., K.K., J.S., A.N., R.A.K. and M.S.; Project administration, A.D.-H. and M.S.; Resources, J.S., A.N., R.A.K and M.S.; Software, K.K., J.S., A.M., A.N. and R.A.K.; Supervision, A.D.-H., K.K., A.N. and R.A.K.; Validation, K.K., J.S., A.M., A.N., R.A.K. and M.S.; Visualization, A.D.-H., A.N. and R.A.K.; Writing - original draft, A.D.-H., K.K., A.M.; final version, A.D.-H.,  A.N. and R.A.K.

\section*{Additional information}

\textbf{Competing interests} The authors declare no conflict of interest. \\
\textbf{Data availability statement:} Data will be available on request. Correspondence and requests for materials should be addressed to A.D.-H.

\end{document}